\documentstyle[12pt,aaspp4,flushrt]{article}

\def\au{{\rm\,AU}}

\def\mearth{{\,M_\oplus}}

\def\10^#1{\times 10^{#1}}

\def\gapprox{\hbox{ \lower0.5ex\hbox{$\sim$} \kern-1.1em 
                 \raise0.5ex\hbox{$>$} }} 
\def\lapprox{\hbox{ \lower0.5ex\hbox{$\sim$} \kern-1.1em 
                 \raise0.5ex\hbox{$<$} }} 

\lefthead{Levison \& Stern}
\righthead{Inclinations in the Kuiper belt}

\begin{document}

\title{On the Size-Dependence of the Inclination Distribution of the Main Kuiper Belt}

\author{Harold F. Levison}

\and

\author{S. Alan Stern}
\affil{Department of Space Studies, Southwest Research Institute,
       Boulder, CO 80302}

\begin{abstract}

We present a new analysis of the currently available orbital elements
for the known Kuiper belt objects.  In the non-resonant, main Kuiper
belt we find a statistically significant relationship between an
object's absolute magnitude ($H$) and its inclination ($i$).  Objects
with $H<6.5$ (i$.$e$.$ radii $\gapprox 170$km for a 4\% albedo) have
higher inclinations than those with $H>6.5$ (radii $\lapprox
170\,$km).  We have shown that this relationship is not caused by any
obvious observational bias.  We argue that the main Kuiper belt
consists of the superposition of two distinct distributions.  One is
dynamically hot with inclinations as large as $\sim 35^\circ$ and
absolute magnitudes as bright as $4.5$; the other is dynamically cold
with $i\lapprox 5^\circ$ and $H>6.5$.  The dynamically cold population
is most likely dynamically primordial.  We speculate on the potential
causes of this relationship.

\end{abstract}



\clearpage
\section{Introduction}
\label{sec_intro}

The discovery of the Kuiper belt in 1992 (Jewitt \& Luu~1993) issued
in a new era for the study of the outer solar system.  The Kuiper belt
is important not only because it is a rich, new region of the solar
system to be explored, but because it contains important fossil clues
about the formation of the outer solar system in particular, and about
planet formation in general.

Since its discovery, the Kuiper belt has supplied us with surprise
after surprise.  For example, before it was discovered, theorists
believed that the Kuiper belt would consist of objects on
low-inclination, nearly-circular orbits beyond the orbit of Neptune
(Levison \& Duncan~1993; Holman \& Wisdom~1993).  This belief seemed
to be confirmed with the discovery of the first two Kuiper Belt
Objects (hereafter KBOs), 1992~QB$_1$ and 1993~FW.  However, the next
four objects discovered revealed a real surprise.  At the time of
discovery their heliocentric distances were close enough to Neptune's
orbit that their orbits should be unstable, unless protected by some
dynamical mechanism.  Indeed, many believed that they might have been
Neptunian Trojans.  However, these were the first discoveries of an
unexpected population of objects on highly eccentric (up to 0.3)
orbits in the 2:3 mean motion resonance with Neptune (co-orbiting with
Pluto).

Currently, objects in the trans-Neptunian region are divided into two
main groups (see Malhotra et al$.$~2000 for a review).  The {\it
Kuiper belt} consists of objects that are primarily on long-lived
orbits, while the {\it scattered disk} consists of objects that have
suffered a close encounter with Neptune (Duncan \& Levison~1997; Luu
et al$.$~1997).  The Kuiper belt itself is typically subdivided into
two populations.  Inside of roughly $42\au$, objects tend to be locked
into mean motion resonances with Neptune.  Most known objects in this
class are in Neptune's 2:3 mean motion resonance.  However, a fraction
also reside in the 3:5 and the 3:4 resonances.  The orbits of all
these objects are probably a result of resonance capture during the
slow outward migration of Neptune during the late stages of planet
formation (Malhotra~1995).

Beyond $42\au$, although several objects are believed to be in the 1:2
mean motion resonance (Marsden~2000a), most objects are not on
resonant orbits.  These non-resonant objects are members of what has
come to be called the {\it main Kuiper belt}.  Models of planetary
migration (e$.$g$.$ Malhotra~1995; Holman~1995; Hahn \& Malhotra~1999)
predict that unlike the KBOs in mean motion resonances, main KBOs
should be on relatively low-inclination, nearly-circular orbits.
However, recent observations have shown that this is not the case.
Numerous objects in this region have very large
inclinations\footnote{Eccentricities are not a good measure of how
excited the Kuiper belt is since most large eccentricity orbits are
removed through close encounters with Neptune, truncating the
eccentricity distribution.  Inclinations do not suffer from this
problem (Duncan, et al$.$~1995).}, certainly up to about
$32^\circ$, and most likely even higher (Marsden~2000a).

Several papers have been published which attempt, among other things,
to explain the high inclinations seen in the main Kuiper belt. The
mechanisms invoked to date involve the scattering of KBOs by large
objects temporarily evolving through the region.  It takes a massive
object to excite KBOs to high inclination; much more massive than the
KBOs themselves\footnote{A simple calculation based on an object's
escape velocity shows that it must be larger than roughly twice the
radius of Pluto to scatter a Kuiper belt object to an inclination of
30$^\circ$.}.  Petit et al$.$~(1999) suggest that the dynamically
excited Kuiper belt is caused by the passage of Earth-mass objects
through that region of the solar system.  Thommes et al$.$~(1999)
suggest that the large inclinations are due to the passage of Uranus
and/or Neptune through the Kuiper belt while on eccentric orbits,
after these planets were ejected from the region between Jupiter and
Saturn.  Ida et al$.$~(2000) suggest that the Kuiper belt was excited
by a passing star.

In this paper we present an analysis of the currently available orbital
data of main belt KBOs which shows a new and surprising trend --- an
unexpected and intriguing correlation between inclination and absolute
magnitude.  In particular intrinsically bright objects tend to be
found on larger inclinations than do intrinsically faint objects.  In
\S{\ref{sec_data}} we present the data and discuss the statistical
significance of this trend.  In \S{\ref{sec_bias}} we investigate
whether this trend is a result of observational selection effects.
Our preliminary interpretation of this trend is presented in
\S{\ref{sec_interp}}. We summarize our findings in
\S{\ref{sec_concl}}.

\clearpage
\section{Observations}
\label{sec_data}

The KBO orbital elements we employ here were taken from the Minor
Planet Center's web site ({\tt
http://cfa-www.harvard.edu/cfa/ps/lists/TNOs.html} for October 20,
2000; Marsden~2000a).  Before we describe our results, however, we
first caution the reader about the use of such data.  Although the
orbital elements in this dataset are given to several significant
figures, many of them are uncertain, and significant changes for
individual objects routinely occur as more data is collected.  This is
particularly severe for objects that have been observed for only one
season (B. Marsden, pers. comm.).  Thus, we restrict our analysis to
objects that have been observed over multiple oppositions.  There are
124 such objects in our dataset; roughly a third of the total.

In general, the inclination, $i$, is the best determined of the 6
orbital elements because it is uniquely determined by the motion of
KBO perpendicular to the ecliptic.  For an object in the ecliptic and
at opposition (where most KBOs have been discovered), observations
taken over even just a short period of time allow for a determination
of its instantaneous heliocentric distance, but do not allow for a
unique determination of the semi-major axis, $a$, or eccentricity,
$e$.  However, since the instantaneous heliocentric distance {\it is}
well determined (being directly calculated from the observed rate of
motion), we do have a good estimate of the object's absolute magnitude
($H$).  \footnote{In planetary science, an absolute magnitude is
defined as the magnitude that an object would have if it were $1\au$
from both the Sun and the Earth and seen at zero phase angle, i$.$e$.$
at opposition.  Such a geometry can never happen in nature, but this
definition is numerically convenient.}

It also should be noted that the MPC dataset suffers from a host of
observational selection effects, including those that affect
inclination.  Surveys for KBOs tend to search near the ecliptic and
thus there is a strong selection against objects with large
inclinations.  Analysis of this and other observational biases is
complicated by the fact that these objects were discovered by many
different observing teams using different equipment and different
search methods.  Thus, the observational biases and limiting
magnitudes vary from object to object.  This complication makes it
difficult to statistically analyze the KBO orbital dataset for trends.
We return to this issue in \S{\ref{sec_bias}}.

Since many objects in mean motion resonances have had their
inclinations affected by these resonances, we restrict ourselves to
objects in the main Kuiper belt.  We define members of the main Kuiper
belt as those objects with $a>42.5\au$ (outside Neptune's 3:5 mean
motion resonance) and $e<0.2$ (to avoid objects in Neptune's 1:2 mean
motion resonance and the scattered disk; Duncan \&
Levison~1997)\footnote{The results presented below are not
significantly sensitive to our choice of these limits.  For example,
if we included objects with $a>41.5\au$ (starting inside of Neptune's
3:5 mean motion resonance) and $e<0.25$, we include 10 more objects in
our sample, but neither our qualitative arguments nor our qualitative
measures of statistical significance change noticeably.}.  There are
80 objects that meet these criteria.  

Figure~\ref{fig_iH} shows the inclinations of these objects as a
function of their absolute magnitude.  The inclinations in this figure
are accurate to better than $\pm 0.5^\circ$, while the absolute
magnitudes are accurate to about $\pm 0.5$ magnitudes (B. Marsden,
pers. comm.).  Notice that this figure indicates a distinct difference
in the character of the inclinations for objects that have $H<6.5$
compared to those with $H>6.5$.  In order to further illustrate this
point, we provide Figure~\ref{fig_isort}, which shows the cumulative
inclination distribution for the two populations.  We refer to the
absolute magnitude boundary between these groups as $H_{break}$.

The natural conclusion from Figures~\ref{fig_iH} and
\ref{fig_isort} is that the inclination distribution of the
intrinsically faint ($H>6.5$) objects appears to be significantly {\it
lower} than the intrinsically bright objects.  Indeed, the median
inclination of the faint objects is $2.2^\circ$, but the median
inclination of the bright objects is $12^\circ$. Of course, assuming
that there is no systematic variation of KBOs albedos, the
intrinsically bright objects represent the largest KBOs \footnote{If
$p=0.04$ then $H=6.5$ implies an object with a radius of $\sim
170\,$km.}. Thus, Figures~\ref{fig_iH} and \ref{fig_isort} suggest
that the largest of the objects in the main Kuiper belt are more
dynamically excited than smaller objects.  This result is surprising
because the mechanisms thus far suggested for exciting the Kuiper belt
(see \S{\ref{sec_intro}}) have predicted such a behavior (however see
Thommes et al$.$~2000).  Because in each of these scenarios the
perturber that excites the Kuiper belt is much larger than the KBOs,
the response of a KBO to the perturber should be virtually independent
of its size.

Before we discuss our interpretation of our new result, we first wish
to demonstrate that this finding is statistically significant.  After
all, there are only 8 objects in our sample with $H<6.5$, so in
principle, small number statistics could be responsible for this
result.  In order to address this issue we employ the
Kolmogorov-Smirnov (K-S) statistical test (Press et al$.$~1992), which
calculates the probability that two distributions are derived from the
same parent distribution, where a zero probability means the
distributions are dissimilar, and unit probability means they are the
same.  We find that the K-S probability of the two inclination
distributions seen in Figure~\ref{fig_isort} is 0.03.  Thus, it is
unlikely that the two distributions are the same\footnote{The data in
Figure~\ref{fig_isort} consists of multiple opposition objects only.
If we repeat this analysis using all 211 main Kuiper belt objects with
both single and multiple opposition orbits, we find a K-S probability
of $0.001$.  Recall that the inclination and absolute magnitude of the
single opposition objects are fairly well known.  The uncertainty is
whether they are members of the main Kuiper belt. Also, if we
broaden our definition of the main Kuiper belt to objects with
$a>41.5\au$ and $e<0.25$, we find a K-S probability of $0.04$.}, and
we can rule out that the two populations are the same at the 97\%
confidence level.

We must also be careful so as to not fortuitously choose a value of
the transition absolute magnitude, $H_{break}$ (set to 6.5 above),
which happens to give a low value of the K-S probability.  So, in
Figure~\ref{fig_ksi} we present the K-S probability as a function of
$H_{break}$.  This figure shows that the K-S probability is small for
all values of $H_{break} < 6.5$, but becomes large for values fainter
than this.  This result can be understood by considering
Figure~\ref{fig_iH}. If $H_{break} < 6.5$, we have only dynamically
hot objects in the bright population, and since one is only adding a
few dynamically hot objects to the faint group, the inclination
distribution of the two groups remain roughly unchanged.  If
$H_{break} > 6.5$, one starts adding dynamically cold faint objects to
the bright group.  Since the cold population far outnumbers the hot
bright population, cold objects start to dominate the bright group as
$H_{break}$ becomes larger than $6.5$.  So, the two distributions look
similar.

In short, Figure~\ref{fig_ksi} shows that our choice of
$H_{break}=6.5$ is not just fortuitous and does not lead us to a false
conclusion about the statistical significance of our finding.  Thus,
we conclude that objects with intrinsic brightnesses greater than
$H_{break}=6.5$ actually do have an inclination distribution that is
statistically different from that of fainter objects.

Could dynamical friction or physical collisions significantly modify
an inclination distribution where the large objects have higher
inclinations?  The response timescale (Binney \& Tremaine~1987) of
large KBOs to dynamical friction in a dynamically cold, ancient Kuiper
belt of $50\mearth$ (see Stern~1996) is $\sim10^9$ years.  However,
after dynamical excitation to eccentricities and inclinations
characteristic of the present-day Kuiper belt, this timescale
increases to $\gapprox 10^{12}$ years.  The lower mass of the Kuiper
Belt which exists today increases this timescale to $\gapprox 10^{14}$
years.  A second potential way of modifying inclinations is
through physical collisions.  However, the time required for a 100-km
class KBO to impact a significant fraction of its own mass in a
$50\mearth$ Kuiper belt is also of order $\sim10^9$ years.  Since
we estimate that both the dynamical and collisional relaxation
timescales are of order 100 times longer than the time required for an
excited, massive KB to erode due to collisions (Stern \& Colwell
1997), one must conclude that the dynamical configuration of the
ancient objects in the present-day, main Kuiper belt is a
well-preserved, fossil remnant of the excitation event(s) itself.

\clearpage
\section{Regarding Potential Observational Biases}
\label{sec_bias}

In this section we investigate whether the differences seen in the
inclination distributions of the bright and faint main Kuiper belt
objects could be the result of observational biases.  As we described
above, this is a difficult issue because these objects were discovered
with a variety of instrumentation and under a variety of observing
conditions.  In particular, the surveys that discovered the faint
objects tend to have limited sky coverage, so they would not have
found the bright objects, which are rare.  On the other hand, the
surveys that covered the most sky have fairly bright limiting
magnitudes, so they would not have discovered the faint objects.  Our
task is made still more difficult because many surveys remain
unpublished, and the details of how these discoveries were made are
unknown.

Here we investigate the only two possible observational selection
effects that we could think of that could erroneously lead us to the
results of the last section.  First, as we described above, the faint
objects tend to be discovered by different surveys than the bright
objects.  The probability of discovering an object of a particular
inclination is a strong function of the ecliptic latitude of the
discovery images.  Images taken at high ecliptic latitude cannot
discover low inclination objects.  On the other hand, images taken at
low ecliptic latitude are biased against discovering high inclination
objects.

The results shown in Figures~\ref{fig_iH} and \ref{fig_isort}
could be a result of differences in the ecliptic latitude of the
discovery images.  For example, if the surveys that covered a large
area of the sky tend to stray further from the ecliptic, we might see
the type of distributions seen in Figures~\ref{fig_iH} and
\ref{fig_isort}.  Figure~\ref{fig_lat} shows the ecliptic latitude of
the objects in our sample at the time of their discovery as a function
of their absolute magnitude.  This data shows that the bright objects
tend to be found at the same ecliptic latitudes as the faint objects.
Indeed, we performed a K-S test similar to that above using
ecliptic latitude instead of inclination and found the K-S probability
is larger than 0.5 for all values of $H_{break}$. Thus, the findings
discussed in \S{\ref{sec_data}} cannot be explained away by discovery
selection effects.

Selection effects on the recovery of objects could also in principle
erroneously lead to the results obtained in \S{\ref{sec_data}}.  It is
well known that the brightest KBOs attract more followup observations
then the faint ones.  This is because the faint objects require large
telescopes on which it is difficult to obtain observing time.  As such
the fainter objects tend to be preferentially lost.  Of the objects in
the main belt discovered before the year 2000 (so there was
opportunity for them to have been observed during a second
opposition), all the objects with $H<5.5$ have been recovered, while
only 36\% of the objects with $H>7.5$ have been observed again.  If,
for the faint objects, there is a selection against recovering high
inclination objects, then the findings of \S{\ref{sec_data}} could be
in error.  To check this possibility, Figure~\ref{fig_rec}
shows the fraction of main belt KBOs fainter than 6.5 that have been
recovered as a function of their inclination.  We only include those
objects that have discovered before the year 2000.  The error-bars
represent the error in the mean; they increase in size with
inclination because there are fewer high inclination objects.  Note
that the recovery fraction for these objects is independent of
inclination. Thus, the finding that objects with $H<6.5$ tend to have
larger inclinations than objects with $H>6.5$ is also {\it not} a
result of recovery statistics.

\clearpage
\section{Interpretation}
\label{sec_interp}

Perhaps the most natural interpretation for the data in
Figure~\ref{fig_iH} is that we are seeing the superposition of two
distinct populations.  The first population contains dynamically hot
objects with inclinations up to $\sim 35^\circ$ and absolute
magnitudes as bright as 4.5. (Of course in the future, members
of this hot population that are larger and/or have higher inclinations
than those currently known, may well be discovered.)  The other
population is a dynamically cold one with $i\lapprox 5^\circ$ and
$H\gapprox6.5$ (radii $\lapprox 170$ for albedo of 4\%).

There are two lines of supporting evidence in our dataset for two
distinct populations.  First, so far in this discussion we have
restricted ourselves to the analysis of inclinations only.  However,
in a dynamically isotropic system, the root-mean-square (RMS) of the
eccentricities should be approximately twice the RMS of the sine of
the inclinations (Lissauer \& Stewart~1993).  So, if our `dynamically
cold' population is real, the eccentricities should also be
small. Indeed, eccentricities should be so small that the eccentricity
distribution of this population should not be truncated by Neptune.
The RMS of the sine of the inclination of objects fainter than $H=6.5$
and with $i\leq 5^\circ$ is $0.039$, which predicts that the RMS
eccentricity should be $0.078$.  It is observed to be $0.076$ which is
in good agreement. The RMS eccentricity of the remaining main belt
objects is $0.11$, which is significantly larger.  Thus, our
dynamically cold population appears to be real.

Our interpretation is also supported by
Figure~\ref{fig_isorts}, which is the same as
Figure~\ref{fig_isort}, but with the $H>6.5$ curve scaled so that the
two curves cross at $i=5^\circ$.  Note that the two distributions are
the same for $i>5^\circ$, arguing that they are members of the same
population.  So, we can conclude from this that the intrinsically
faint objects with $i>5^\circ$ are part of the same population as the
intrinsically bright objects.  If this interpretation is correct, then
approximately 40\% of the objects in our sample are part of the
dynamically excited population.

As we were preparing this manuscript, two papers became available that
also argue for two populations in the main Kuiper belt.  Brown~(2000)
performed detailed modeling of the one-dimensional inclination
distribution of the main Kuiper belt.  Although his results are
somewhat model dependent, owing to an assumed functional form for
the intrinsic inclination distribution of
$\sin{(i)}\exp{(-i^2/2\sigma^2)}$, he concludes that the main Kuiper
belt is most likely composed of the superposition of two distinct
populations --- one dynamically hot and the other dynamically cold.
The dynamically cold population is best fit by $\sigma = 2.2^\circ$,
which is consistent with our estimate that the maximum inclination of
this population is roughly $5^\circ$.

More convincing and relevant, however, are the recent results of
Tegler \& Romanishin~(2000), who have studied the colors of KBOs.  It
has been previously shown that the Kuiper belt and scattered disk most
likely contain two distinct color populations --- one that is
comprised of objects that are gray in color and one in which the
objects are red (Tegler \& Romanishin~1998).  Tegler \&
Romanishin~(2000) found that in the main Kuiper belt, all objects on
low-inclination, nearly-circular orbits are red in color, while the
rest of the KBOs are a mixture of both red and gray colors (also see
Marsden~2000b). The black and red dots in Figure~\ref{fig_iH}
represent those objects for which Tegler \& Romanishin measured a gray
and red color, respectively.  Tegler \& Romanishin's result seems to
indicate that at least the surfaces of the dynamically cold main
Kuiper belt objects are chemically distinct as a group from the rest
of the KBOs.

Based on the various lines of evidence we conclude that the main
Kuiper belt is a superposition of two distinct populations and that
these populations consist of objects with different sizes, different
dynamics, and different surface properties.  We speculate that a
natural explanation for this result is as follows\footnote{In the
following scenario we are assuming that the differences in $H$ are due
to differences in size.  It is possible, but less likely, that it
could be due to albedo differences.}.

Initially the protoplanetary disk in the Uranus-Neptune region
and beyond was dynamically cold with size distribution and color that
varied with heliocentric distance.  In particular, significant numbers
of large objects ($H<6.5$) had only formed in the inner regions of the
disk while few, if any, objects this large formed in the outer
regions.  Then a dynamically violent event cleared the inner region
of the disk, dynamically scattering the inner-disk objects outward.
Most of these objects were either ejected from the solar system,
placed in the Oort cloud, or became members of the scattered disk.
However, a few of these objects would have been deposited in the main
Kuiper belt, becoming the dynamically hot population described above.

This scenario has several implications.  First, it suggests that 
objects in the scattered disk, the dynamically hot main Kuiper belt, and
perhaps in Neptune's mean motion resonances should have similar
size-distributions and physical characteristics because they were
all populated with the objects initially in the inner disk.  In addition,
since current models of the Kuiper belt show that the cold population
is likely to be dynamically stable (Duncan et al$.$~1995), this
population should not be contributing significantly to the Centaurs.
Hence, the Centaurs should also have a size-distribution and physical
properties similar to the dynamically hot main Kuiper belt and its
cohorts.  This appears to be born out by observations.  Tegler \&
Romanishin~(2000) find that the scattered disk, the dynamically hot main
Kuiper belt, the plutinos, and the Centaurs roughly have the same mixture
of red and gray objects.  In addition, all these regions contain
objects with $H<6.5$.

Our scenario also suggests that the dynamically cold population is a
dynamically primordial population; member objects most likely formed
near where they are observed and have not been significantly perturbed
over the age of the solar system.\footnote{By member objects, we
specifically exclude recently created collisional shards (see e.g.,
Farinella et al.~2000).}  It also suggests that because the
intrinsically brightest objects in this population have $H \sim 6.5$ and
other brighter (larger) objects have been found in the main Kuiper belt,
that the largest object to grow in this region has $H=6.5$ or a radius
of $\sim 170\,$ km (4\% albedo).  This result may supply
important constraints on the accretional history of this region,
possibly including coonstraints on the solid surface density of material
in the region and the date of the event(s) that dynamically excited the
Kuiper belt.

\clearpage
\section{Summary}
\label{sec_concl}

We have shown that the inclination distribution of objects in the main
Kuiper belt most likely varies as a function of absolute magnitude.
In particular, objects intrinsically brighter than $H=6.5$ appear to
have systematically higher inclinations than intrinsically fainter
objects.  There is only $\sim 3\%$ chance that these two distributions
are the same.  We have shown that this result is unlikely to be caused
by biases in discovery or recovery observing procedures.  Therefore,
although it is possible that this conclusion is a result of small
number statistics, we believe that it is real.  Future discoveries and
followups will clearly resolve this issue.  The clear implication of
our result is that a main belt object's inclination is dependent on
its size.

The differences between intrinsically bright objects and the
intrinsically faint objects is best seen in Figure~\ref{fig_iH}.
Perhaps the most natural interpretation for the data in this figure is
that we are seeing the superposition of two distinct populations.  The
first contains a dynamically hot population (inclinations up to $\sim
35^\circ$) consisting of both large and small objects (absolute
magnitudes as small as 4.5 or radii up to $\sim 330\,$km for albedos
of 4\%).  Indeed, even larger objects and/or objects with higher
inclinations are likely to still be found.  The other population is a
dynamically cold one ($i\lapprox 5^\circ$) preferentially containing
smaller objects ($H\gapprox6.5$ or radii $\lapprox 170\,$km for
albedos of 4\%).

\acknowledgments

We would like to thank L. Dones, B. Gladman, and P. Tamblyn for useful
discussions.  We are also grateful to W. Bottke, R. Canup, M. Duncan
and an anonymous referee for comments on an early version of
this manuscript.  We also thank NASA's PGG and Origins programs for
support.

\clearpage

\figcaption[] {\label{fig_iH} The inclination ($i$), absolute
magnitude ($H$) distribution of multiple opposition objects in the
main Kuiper belt as of October 20, 2000.  Note that objects brighter
than $H=6.5$ are dynamically more excited than those with $H>6.5$.
The red dots represent red objects for which Tegler \&
Romanishin~(2000) measured a $V$-$R >0.6$. The black dots represent
gray objects for which they measured a $V$-$R <0.6$.  The blue dots
represent objects for which they have not measured colors.}

\figcaption[] {\label{fig_isort} The cumulative inclination
distribution for members of the main Kuiper belt with multiple
opposition orbits.  The population is divided into two groups.  The
solid curve shows only those objects fainter than $H=6.5$, while the
dotted curve only includes objects brighter then this.  A K-S test puts
the probability that these two distributions are the same at $0.03$.}

\figcaption[] {\label{fig_ksi} The K-S probability that the
inclination distribution of objects brighter than $H_{break}$ is the
same as that of objects less than $H_{break}$.  The K-S probability is
small for $H_{break}<6.5$ indicating that the two distributions are
indeed most likely different.}

\figcaption[] {\label{fig_lat} The ecliptic latitude, absolute
magnitude ($H$) distribution of multiple opposition objects in the
main Kuiper belt as of October 20, 2000.  The ecliptic latitude was
calculated at the time of discovery.  Note that there is not a
significant correlation between these two parameters.}

\figcaption[] {\label{fig_rec} The fraction of main belt $H>6.5$ KBOs
that have so far been recovered as a function of their inclination.
We only include those objects that have had the potential for being
observed on multiple oppositions.  The error-bars represent the error
in the mean.}

\figcaption[] {\label{fig_isorts} The same as
Figure~\ref{fig_isort} except that the $H>6.5$ curve is scaled so that
the two curves cross at $i=5^\circ$.}

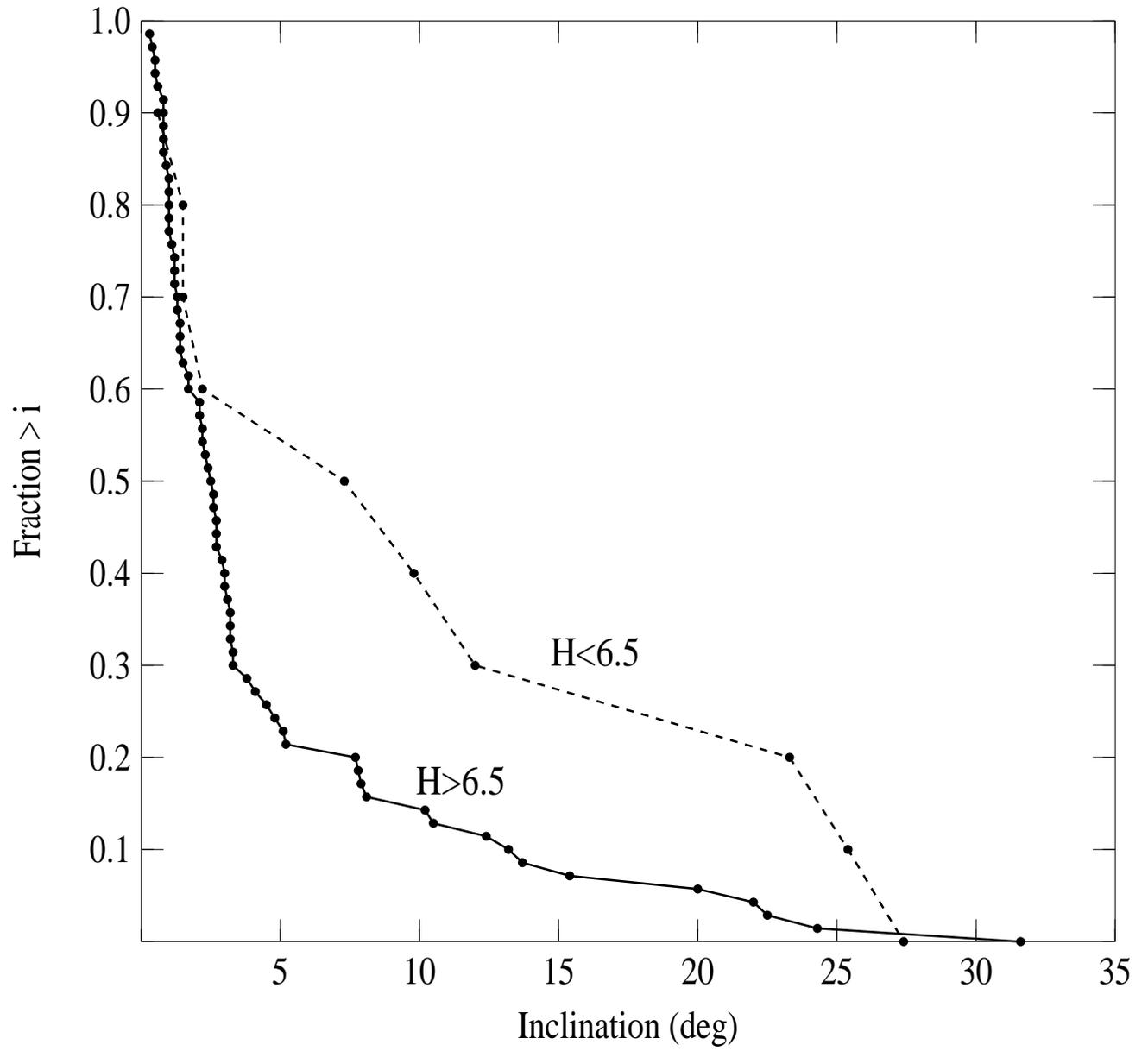
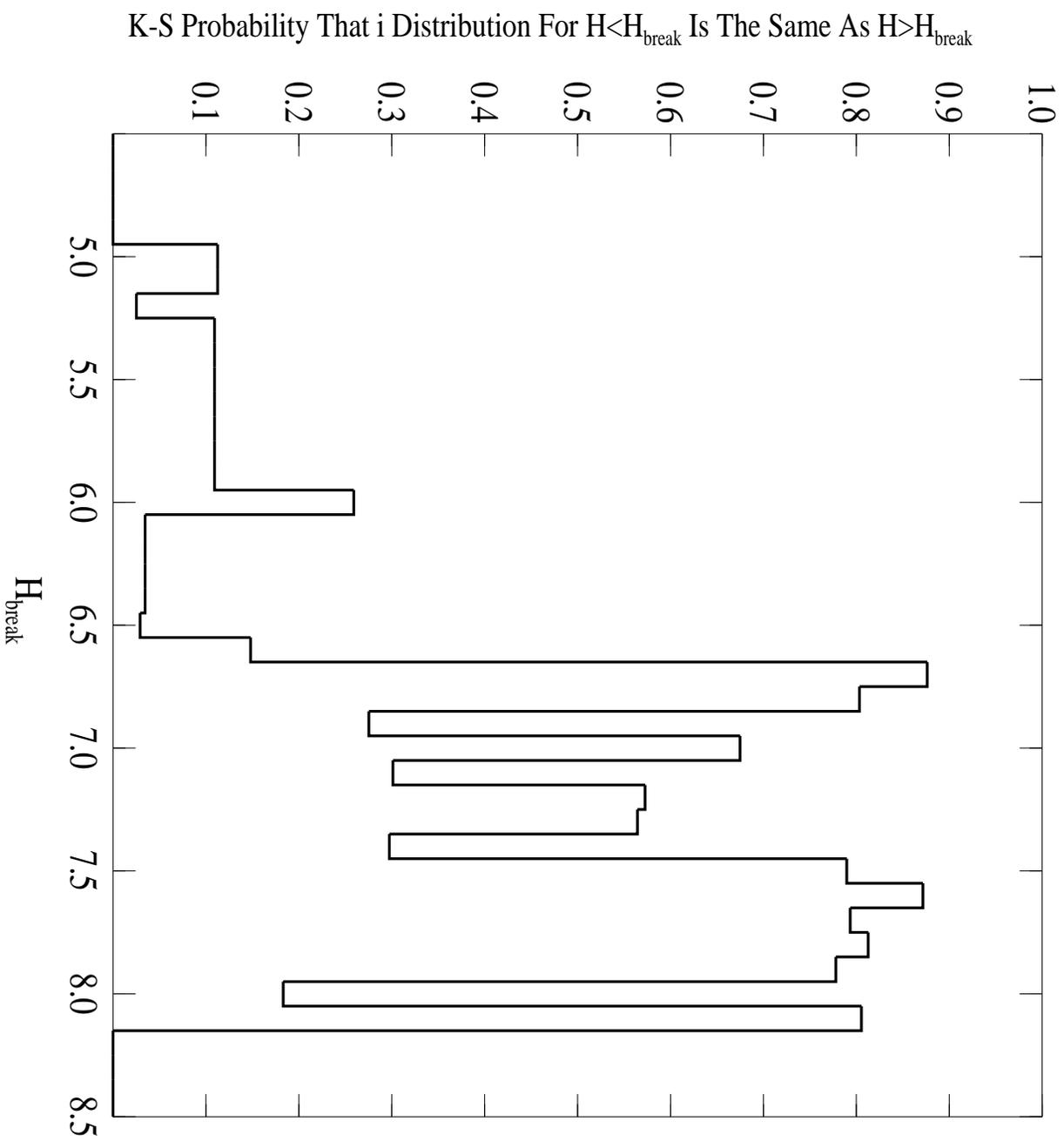
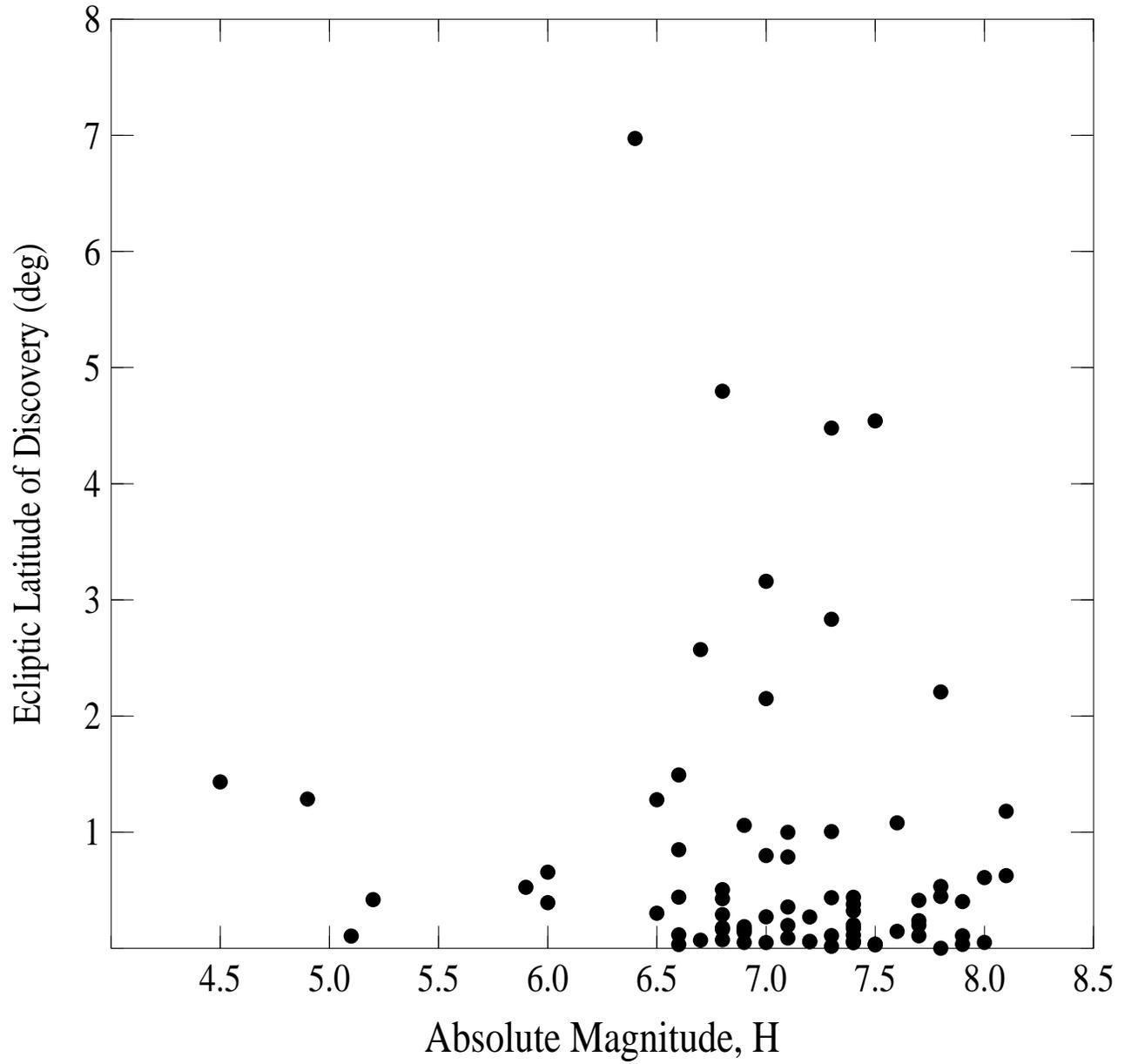
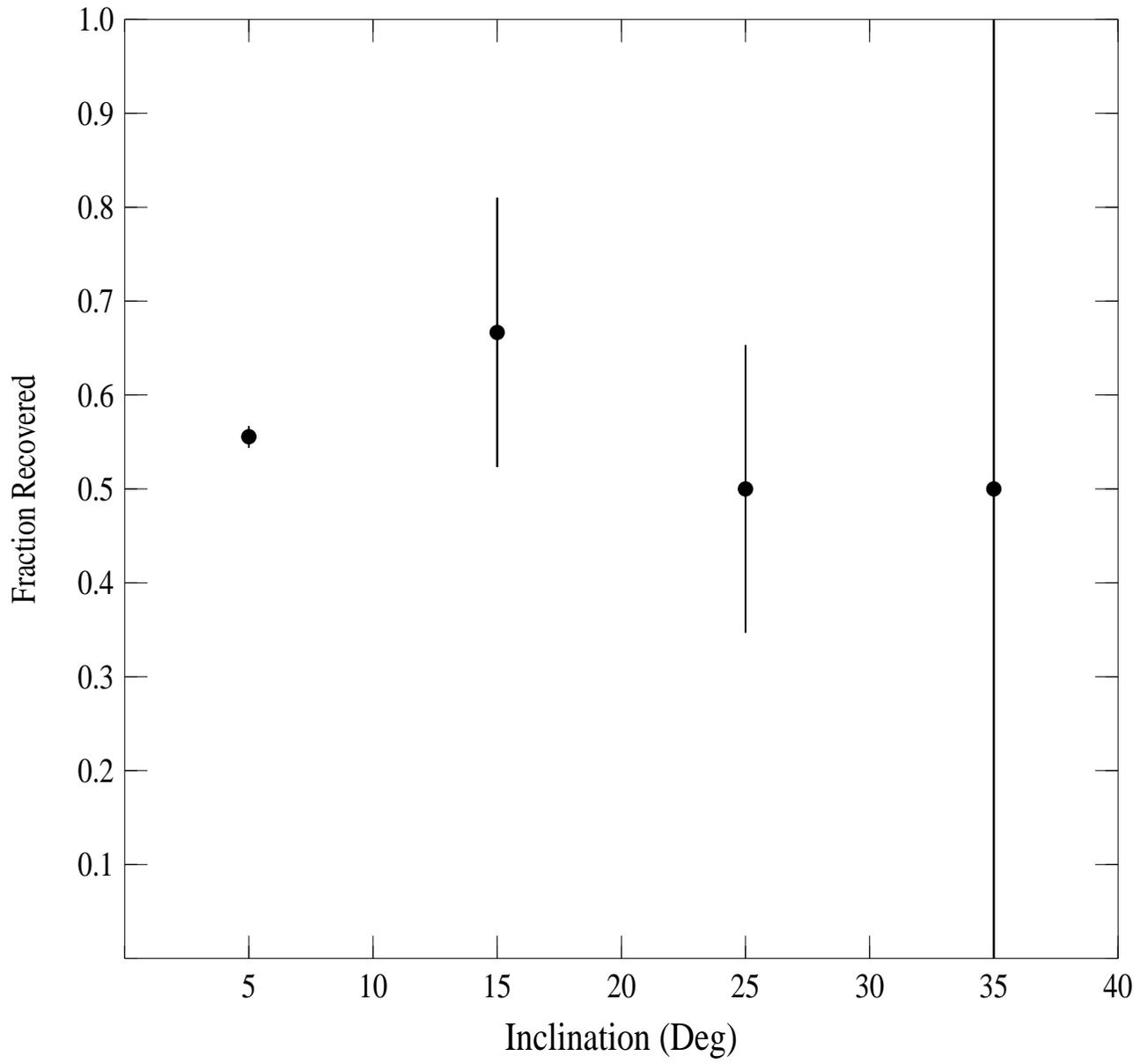
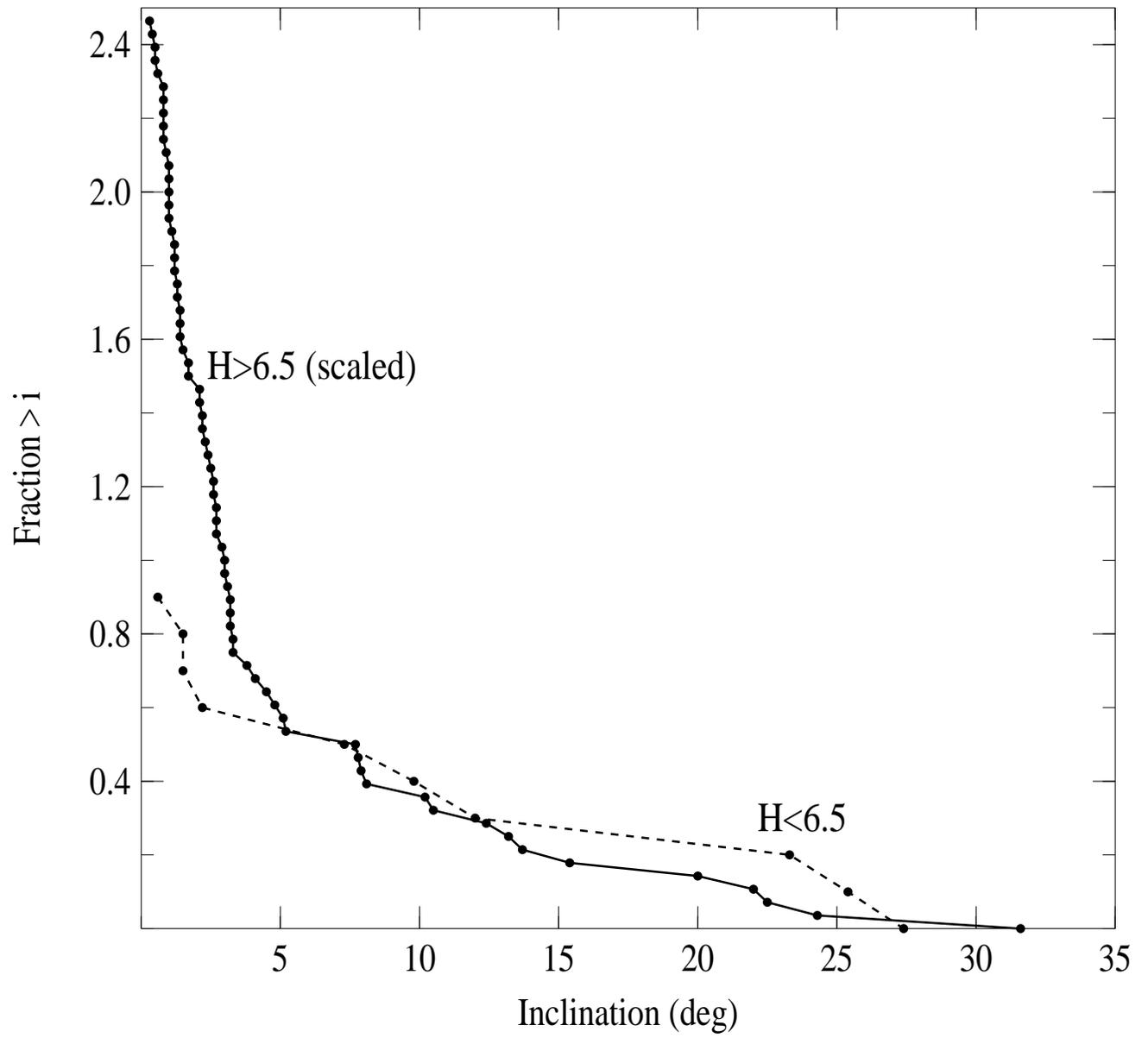
\end{document}